\documentclass[preprint,showpacs,aps,amsmath,amssymb,nofootinbib,superscriptaddress,showkeys]{revtex4}

\usepackage{epsfig}

\def\be{\begin{equation}}
\def\ee{\end{equation}}
\def\bea{\begin{eqnarray}}
\def\eea{\end{eqnarray}}
\def\bes{\begin{subequations}}
\def\ees{\end{subequations}}

\def\beq{\begin{equation}}
\def\eeq{\end{equation}}
\def\barr{\begin{eqnarray}}
\def\earr{\end{eqnarray}}

\def\bb{\alpha_2}
\def\cc{\alpha_3}

\def\lsim{\:\raisebox{-1.1ex}{$\stackrel{\textstyle<}{\sim}$}\:}
\def\gsim{\:\raisebox{-1.1ex}{$\stackrel{\textstyle>}{\sim}$}\:}

\begin{document}

\title[]{Radiatively broken symmetries of nonhierarchical neutrinos}


\author{Amol Dighe}
\affiliation{ Tata Institute of Fundamental Research, 
Homi Bhabha Road, Mumbai 400 005, India}
\author{Srubabati Goswami}
\affiliation{ Harish-Chandra Research Institute, Chhatnag Road, 
Jhusi, Allahabad 211 019, India}
\author{Probir Roy}
\affiliation{ Tata Institute of Fundamental Research, 
Homi Bhabha Road, Mumbai 400 005, India}

\begin{abstract}
Symmetry-based ideas, such as the quark-lepton complementarity (QLC) 
principle and the tri-bimaximal mixing (TBM) scheme, have been  
proposed to explain the observed mixing pattern of neutrinos. We 
argue that such symmetry relations need to be imposed at a high scale 
$\Lambda \sim 10^{12}$ GeV characterizing the large masses of right-handed 
neutrinos required to implement the seesaw mechanism. For nonhierarchical 
neutrinos, renormalisation group evolution down to a laboratory energy 
scale $\lambda \sim 10^3$ GeV tends to radiatively break these symmetries 
at a significant level and spoil the mixing pattern predicted by them. 
However, for Majorana neutrinos, suitable constraints on the extra phases 
$\alpha_{2,3}$ enable the retention of those high scale mixing patterns
at laboratory energies. We examine this issue within the Minimal  
Supersymmetric Standard Model (MSSM) and demonstrate the fact posited above
for two versions of QLC and two versions of TBM. The appropriate constraints 
are worked out for all these four cases. Specifically, a preference for  
$\alpha_2 \approx \pi$ (i.e. $m_1 \approx -m_2$) emerges in each case. We
also show how a future accurate measurement of $\theta_{13}$ may enable 
some 
discrimination among these four cases in spite of renormalization group 
evolution.
\end{abstract}

\pacs{11.10.Hi, 12.15.Ff, 14.60.Pq}

\keywords{neutrino masses and mixing, renormalisation group running,
quark-lepton complementarity, tribimaximal mixing}



\maketitle

\section{Introduction}
\label{intro}

Outstanding recent experiments have increased our knowledge 
\cite{whitepaper} of neutrino masses and mixing angles enormously.
We are already certain that at least two of the three known neutrinos are
massive, the heavier and the lighter of them being
respectively $\gsim 0.05$ eV and $\gsim 0.009$ eV in mass.
We also know that two of the three neutrino mixing angles 
are large: $\theta_{23} \approx 45^\circ$ and $\theta_{12}
\approx 34^\circ$, while the third is significantly smaller:
$\theta_{13} < 12 ^\circ$.
The total sum of the neutrino masses is also cosmologically
bounded from above by ${\cal O}(1)$ eV.
Much remains to be known, though. 
The values of $\theta_{13}$ and the leptonic CP violating
Dirac phase $\delta_\ell$, are still unknown.
So is the ordering of the neutrino masses 
$m_i$ ($i=1,2,3$) -- whether it is
normal ($|m_3| > |m_{1,2}|$) or inverted  ($|m_3| < |m_{1,2}|$).
We also do not know if the three neutrinos  are
hierarchically spaced in mass like charged fermions or if they are
nonhierarchical. Our term nonhierarchical here includes both the inverted 
hierarchical (IH) case , i.e $|m_3| \ll |m_1| \sim |m_2| \sim 0.05$ eV,  
and the quasi-degenerate (QD) situation \cite{qdn}, 
i.e. $|m_1| \sim |m_2| \sim |m_3| \gg 0.05$ eV, the latter with either a normal or an inverted mass ordering. 
Neither of these scenarios is observationally excluded as yet and we focus on them.
As per our present knowledge, the average neutrino mass could still in fact be 
anywhere between half of the atmospheric oscillation mass scale,
i.e. $\approx 0.025$ eV and a third of the cosmological 
upper bound, i.e. $\approx 0.3$ eV.
Finally, most theoretical ideas expect the three neutrinos
to be Majorana particles whose masses $m_i$ can be complex.
In that case, since one of their phases can be rotated away, there are two 
additional, possibly nonzero, 
phases \cite{valle} on which we do not have any direct information
at present.
This is because no convincing evidence exists as yet of 
neutrinoless nuclear double beta decay which is the only known 
direct probe \cite{doublebeta} on these phases.
Indirectly, of course, some constraints on these phases may also arise from
considerations of leptogenesis \cite{blanchet}.
It is nevertheless worthwhile to try to constrain
these phases in some other way. 
That is one of the aims of the present work, which is an elaboration
of our earlier shorter communication \cite{rgqlc} 
with many additional results.
In particular, we demonstrate here that, given the constraints on these
Majorana phases, a measurement of $\theta_{13}$  can 
make some discrimination
among four scenarios considered by us despite renormalization group (RG)
running.

The observed bilarge pattern of neutrino mixing has led to
the idea of some kind of a symmetry at work.
Several symmetry-based relations\footnote{
Here one should perhaps make a distinction between
a symmetry of the Lagrangian and just a special relation
among coupling strengths or masses.
Nevertheless, the relations of concern to us can be
implemented through specific symmetries of the Lagrangian.
}
have in fact been proposed,
which give rise to specific neutrino mixing patterns.
Two of the most promising mixing patterns, that we will 
be concerned with here, are 
(i) quark-lepton complementarity (QLC) 
\cite{qlc0,raidal,minakata,qlc+,ferrandis} and 
(ii) tribimaximal mixing (TBM) \cite{tbm}.
QLC involves bimaximal mixing \cite{bimax}
followed by the unitary transformation of quark mixing.
A bimaximal mixing can in turn be generated by a
$\mu$-$\tau$ exchange symmetry \cite{mutauex},
an $L_\mu-L_\tau$ gauge symmetry \cite{lmu-ltau},
or an $S_3$ permutation symmetry \cite{moha-nussinov}.
The second step is inspired by SU(5) or SO(10) GUT,
as discussed later.
Similarly, 
a tribimaximal mixing pattern may be obtained from an
$A_4$ \cite{a4} or $S_3$ \cite{s3} family symmetry.
However, a major issue in connection with such symmetries is 
the scale at which they are to be implemented.
Neutrino masses and mixing angles are related directly to
the corresponding Yukawa coupling strengths which run with the energy
scale.
There is as yet no universally accepted explanation of
the origin of neutrino masses, but the seesaw mechanism
\cite{seesaw} is the most believable candidate so far.
The form of the light neutrino mass matrix in family space
in that case is ${\cal M}_\nu = - (m_\nu^D)^T M_R^{-1} m_\nu^D$,
where $m_\nu^D$ is the Dirac neutrino mass matrix (analogous
to the charged fermion ones) and $M_R$ the mass matrix for
very heavy right chiral singlet neutrinos. 
If the Dirac mass of the heaviest neutrino is taken to be
1 -- 100 GeV, the atmospheric neutrino data require
typical eigenvalues of $M_R$ to be 
in the $10^{11}$--$10^{15}$ GeV
range \cite{seesawscale}. 
This is also the desirable magnitude for $M_R$ from the
standpoint of a successful leptogenesis \cite{leptogenesis}.
From these considerations, we choose to implement the 
above mentioned symmetries at the scale 
$\Lambda \sim 10^{12}$ GeV. 
One can take issue with the particular value chosen for 
$\Lambda$. However, our conclusions are only logarithmically
sensitive to the precise value of this scale.

A question arises immediately on the application of such a high
scale symmetry on the elements of the neutrino mass matrix.
It concerns their radiative breaking via RG evolution
down to a laboratory energy scale $\lambda \sim 10^3$ GeV.
The actual evolution \cite{rgevol,casas,antusch-majorana}
needs to be worked out in a specific theory
which we choose to be the minimal supersymmetric standard
model (MSSM \cite{mssm}).
That is why we have taken $\lambda$ to be of the order of the explicit 
supersymmetry breaking or the intra-supermultiplet
splitting scale ${\cal O}$(TeV). Once again, our calculations are only logarithmically sensitive
to this exact choice. The point, 
however, is that -- for nonhierarchical neutrinos -- 
symmetry relations formulated at $\Lambda$ will in general get spoilt on
evolution down to $\lambda$.

The full RG equations for the evolution of neutrino masses and mixing
angles in the MSSM have been worked out 
\cite{casas,antusch-majorana} in detail. 
In particular, the evolution effects on the mixing angles
are found to be controlled by the quantities \cite{rgqlc}
$\Delta_\tau |m_i+m_j|^2 / (|m_i|^2 - |m_j|^2)$
where $\Delta_\tau$, to be specified later, is a small fraction $\lsim 10^{-2}$, 
while $i,j$ refer to the concerned neutrino mass eigenstates. 
Consequently, these effects are negligible for a normal
hierarchical mass pattern with $|m_3| \gg |m_2| \gg |m_1|$.
RG effects can become significantly large only when
neutrinos are nonhierarchical. 
There is another important characteristic of the 
above-mentioned ratios. 
While their denominators involve only the absolute masses $|m_i|$,
the numerators involve the combinations $|m_i+m_j|^2$.
Therefore, with appropriate constraints on the neutrino 
Majorana phases, the desired symmetry relations can be
approximately preserved at the laboratory scale $\lambda$ even for 
nonhierarchical neutrinos -- in agreement with
the mixing pattern that has emerged from the oscillation data.
The constraints on the majorana phases and the consequent
discrimination among the scenaios by a measurement of
$\theta_{13}$ constitute our main results. 
Our work is somewhat complementary to that of Ref.~\cite{Schmidt} 
in the QLC sector and Ref.~\cite{xing} in the TBM sector.

In this paper we work out in detail the last-mentioned constraints 
on the neutrino Majorana phases in the (i) bimaximal mixing + QLC
and 
(ii) TBM scenarios respectively. Each of these comes in two variations. 
So we have in all four cases at hand. 
Thus, the scope of the present work is much larger than
our earlier shorter communication \cite{rgqlc} which addressed
only one version of QLC and did not consider the 
implications for $\theta_{13}$.
A major technical observation utilized by us is the following.
Suppose $\theta_{13}^\Lambda$, the high scale value of 
the angle $\theta_{13}$, 
is sufficiently small (as is the case for the situations 
considered here) such that 
${\cal O}(\theta_{13})$ terms can be neglected in comparison 
with other ${\cal O}(1)$ 
terms in the RG equations \cite{antusch-majorana}.
Then the neutrino mass matrix ${\cal M}_\nu^\lambda$, 
at the laboratory scale $\lambda$, becomes analytically 
tractable in terms of its high scale form ${\cal M}_\nu^\Lambda$.
In fact, the relation obtained looks quite simple and transparent. 
The step from there to explicit constraints on the neutrino
Majorana phases is then shown to be quite straightforward.
The rest of this paper is organised as follows.
Sec.~\ref{param} contains a description of the
parametrisation that we find convenient to adopt for
nonhierarchical neutrino masses.
In Sec.~\ref{highscale}, we introduce 
two versions each of the QLC and TBM
scenarios to be implemented at the high scale.
In sec.~\ref{evol}, we discuss the 
energywise downward evolution of the 
Pontecorvo-Maki-Nakagawa-Sakata (PMNS) mixing matrix
in general, and its effects on the predictions
of the scenarios under consideration.
In sec.~\ref{constraints} we study the constraints on
neutrino and MSSM parameters in order for the scanarios to
be valid and explore if these scenarios may be distinguished
by means of more accurate measurements of the neutrino mixing angles.
The concluding sec.~\ref{concl} consists of a summary
and the discussion of our main results.

\section{Parametrisation of nonhierarchical neutrino masses}
\label{param}

We work in the convention \cite{pdg} in which the neutrino
mass eigenstates $|\nu_1\rangle, |\nu_2\rangle, |\nu_3\rangle$
are related to the flavour eigenstates
$|\nu_e\rangle, |\nu_\mu\rangle, |\nu_\tau\rangle$
with the unitary mixing matrix $U_\nu$:
\beq
|\nu_\alpha\rangle = U_{\alpha i} |\nu_i\rangle \;,
\eeq
$\alpha$ and $i$ being flavour and mass indices respectively.
We take the neutrino mass term in the Lagrangian to be
\beq
{\cal L}^\nu_{\rm mass} = -\frac{1}{2}
\overline{\nu_{L \alpha}^C} {\cal M}_{\nu \alpha \beta} \nu_{L \beta} + h.c.
\label{l-mass}
\eeq 
Thus,
\beq
U_\nu ^\dagger {\cal M}_\nu U^*_\nu = 
\left( \begin{array}{ccc}
m_1 & 0 & 0 \\
0 & m_2 & 0 \\
0 & 0 & m_3 \\
\end{array} \right) \; ,
\label{diagon}
\eeq
where $m_i$ are in general complex.
However, one of the three phases of
$m_{1,2,3}$ can be absorbed in the overall phase choice of
$\nu_L$ in (\ref{l-mass}).
We can therefore choose
\beq
m_1 = |m_1| ~,~ m_2 = |m_2| e^{i \alpha_2} 
~,~ m_3 = |m_3| e^{i \alpha_3} \; ,
\eeq
where $\alpha_{2,3}$ are real. 
Experiments with atmospheric neutrinos tell us that 
\cite{pdg,nu-fits}
\beq
|\delta m^2_A| \equiv | |m_3|^2 - |m_{2,1}|^2| =
(2.4 \pm 0.3) \times 10^{-3} \mbox{ \rm eV}^2 \;,
\label{dma-value}
\eeq
while experiments with solar electron neutrinos and reactor 
electron antineutrinos yield \cite{pdg,nu-fits}
\beq
\delta m^2_S \equiv |m_2|^2 - |m_1|^2 =
(7.9 \pm 0.4) \times 10^{-5} \mbox{ \rm eV}^2 \;.
\label{dms-value}
\eeq
 
For charged fermions $(f = u, d, l)$, the mass term is
\beq
{\cal L}_{\rm mass}^f = -\frac{1}{2}
\overline{f _{R \alpha}} m_{f \alpha \beta} f_{L\beta} + h.c.
\eeq
The corresponding mass matrix $m_f$ is put into a diagonal form by
\beq
U_f^\dagger m_f^\dagger m_f U_f = |m_f^{(D)}|^2 \; .
\eeq
Now the unitary Cabbibo-Kobayashi-Maskawa (CKM)
and the Pontecorvo-Maki-Nakagawa-Sakata (PMNS) mixing matrices,
whose elements contribute to the observed quark and neutrino
processes respectively, are given by
\bea
V_{\rm CKM} & = & U^\dagger_u U_d \; ,
\label{vckm-def} \nonumber \\
U_{\rm PMNS} & = & U^\dagger_\ell U_\nu 
\label{upmns-def}
\; .
\eea 
One can write $U_{\rm PMNS}$ in the standard basis \cite{pdg}
in terms of the angles $\theta_{12}, \theta_{23}, \theta_{13}$
and the CP violating phase $\delta_\ell$.
With $s_{ij} \equiv \sin \theta_{ij}$ and
$c_{ij} \equiv \cos \theta_{ij}$,
\beq
U_{\rm PMNS} = \left( \begin{array}{ccc}
c_{12}c_{13} & s_{12}c_{13} & s_{13}e^{-i\delta_\ell} \\
-s_{12}c_{23}-c_{12}s_{23}s_{13}e^{i\delta_\ell}  
& c_{12}c_{23}-s_{12}s_{23}s_{13}e^{i\delta_\ell} & s_{23}c_{13} \\
s_{12}s_{23}-c_{12}c_{23}s_{13}e^{i\delta_\ell} & 
-c_{12}s_{23}-s_{12}c_{23}s_{13}e^{i\delta_\ell} & c_{23}c_{13} \\
\end{array} \right) \; .
\eeq
The experiments mentioned earlier then also tell us that 
\cite{pdg,nu-fits}
\bea
 \theta_{12} & = & 33.9^\circ \pm 1.6^\circ \; ,\nonumber \\
\theta_{23} & = & 43.3^\circ \pm 8.2^\circ \; , \nonumber \\
\theta_{13} & < & 12^\circ \; .
\eea

We find it convenient to parametrise the absolute masses $|m_i|$
for nonhierarchical neutrinos in terms of  three real
parameters $m_0, \rho_A$ and $\epsilon_S$ as follows: 
\bea 
|m_1| &=& m_0 (1 - \rho_A) (1 - \epsilon_S) \; , \nonumber \\
 |m_2| &=&m_0 (1 - \rho_A) (1 + \epsilon_S) \; , \nonumber \\
 |m_3| &=& m_0 (1 + \rho_A) \; .
\label{rhoeps-def}
\eea 
In eqs.~(\ref{rhoeps-def}), $m_0$ defines the overall mass scale 
of the neutrinos, whereas $\rho_A$ and $\epsilon_S$ are
dimensionless fractions with $-1 \leq \rho_A \ll 1$ and
$0 < \epsilon_S < |\rho_A|$ for nonhierarchical neutrinos. 
The sign of $\rho_A$ is positive (negative) for a normal (inverted) ordering 
of neutrino masses. Moreover, $\rho_A \approx -1$ ($|\rho_A| \ll 1$)
for the IH (QD) case; in either case $\epsilon_S \ll 1$.
For comparison, it may be noted that for normally hierarchical neutrinos, 
$\rho_A \sim \epsilon_S \sim 1$. 
We can further write the solar and atmospheric neutrino mass
squared differences as 
\bea
\label{dmsq-rhoeps}
\delta m^2_S &=& |m_2|^2 - |m_1|^2 \approx 4m^2_0 (1 - \rho_A)^2 
\epsilon_S \; ,
\phantom{+ {\cal O}(\epsilon^2_S)} \nonumber \\
|\delta m^2_A| &=& ||m_3|^2 - (|m_1|/2 + |m_2|/2)^2 | = 
4 m^2_0 |\rho_A| \; .
\eea
Utilizing (\ref{dms-value}), (\ref{dma-value}) and 
(\ref{dmsq-rhoeps}), we see that 
\beq
m_0 > 0.024 ~{\rm eV} \;.
\eeq
Also, the cosmologically bounded sum of neutrino absolute
masses is given by
\beq
\Sigma_i |m_i|  =  3 \, m_0 (1 - \rho_A/3) \lsim 1 ~{\rm eV} \; .
\label{sigma-mi}
\eeq
From (\ref{dmsq-rhoeps}) and (\ref{sigma-mi}), it follows that
\beq
\frac{4}{9} \frac{(\sum |m_i|)^2}{|\delta m^2_A|} =
\frac{(1 - \rho_A/3)^2}{|\rho_A|} \; .
\label{4/9form}
\eeq
Utilising (\ref{dma-value}), (\ref{4/9form}) 
and the cosmological upper bound  
(\ref{sigma-mi}), we get $|\rho_A| \gsim 5.5 \times 10^{-3}$.

\begin{figure}[t]
\epsfig{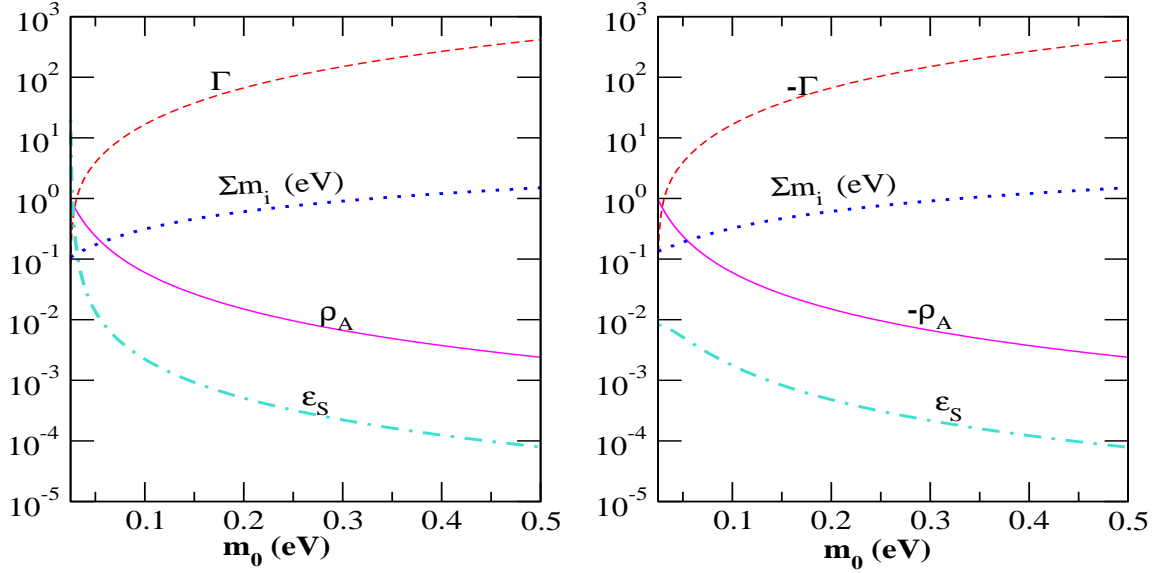}
\caption{The parameters $\rho_A$, $\epsilon_S$ and $\Gamma$
as well as the value of $\sum |m_i|$ (eV) as  functions of
$m_0$, for a normal (left panel) and an inverted (right panel)
mass ordering of neutrinos.
Note that $\rho_A$ and $\Gamma$ are negative for 
inverted mass ordering.
For normal ordering, $\epsilon_S$ goes to near unity
with low $m_0$. 
\label{plot-rhoeps}}
\end{figure}

In Fig.\ref{plot-rhoeps}, we show how $\rho_A$, $\epsilon_S$
and $\sum |m_i|$ behave as functions of $m_0$. We also find
it convenient to define the derived dimensionless parameter
\beq
\Gamma \equiv \frac{1}{\rho_A} - \rho_A \; ,
\eeq
whose behaviour is included in the figure.
Since $|\rho_A| < 1$, the sign of $\Gamma$ is 
positive ($\rho_A > 0$) for a normal ordering and 
negative ($\rho_A < 0$) for an inverted ordering of the
neutrino masses.
The inequality $5.5 \times 10^{-3} \lsim |\rho_A| < 1$
translates to $|\Gamma| \lsim 180$.
We can generate typical numbers for nonhierarchical neutrinos: 
In the IH scenario, for instance, if $m_0$ is chosen as
0.025 eV, we get $\rho_A \approx -1$, $\Gamma \approx 0^-$, and 
$\epsilon_S \approx 8 \times 10^{-3}$.
On the other hand, for QD neutrinos, choosing $m_0 = 0.2$ eV,
we have $|\rho_A| \approx 1.5 \times 10^{-2}$,
$|\Gamma| \approx 65$,
and $\epsilon_S \approx 5 \times 10^{-4}$.

\section{Symmetries and mixing angles at the high scale}
\label{highscale}

The neutrino mass matrix at the high scale $\Lambda$ originates from
a dimension-5 operator
\beq
{\cal O} = c_{\alpha \beta} \frac{(\ell_\alpha . H)
(\ell_\beta . H)}{\Lambda} \; .
\label{operator}
\eeq
In (\ref{operator}), $\ell_\alpha$ and $H$ are the SU(2) doublet 
lepton and Higgs fields respectively and $c_{\alpha \beta}$
are dimensionless coefficients that run with the energy scale. Then
\beq
{\cal M}^\Lambda_{\nu \alpha \beta} \sim
c_{\alpha \beta} \frac{v^2}{\Lambda} \;,
\eeq
where $v = 246$ GeV and $\Lambda \sim M_{\rm MAJ}$, 
the Majorana mass characterising the 
heavy SM-singlet Majorana neutrino N. 
The symmetries at the high scale $\Lambda$ give rise to
specific structures for the matrix ${\cal M}_\nu^\Lambda$,
and hence predict the values of the mixing angles 
$\theta_{ij}^\Lambda$ at the scale $\Lambda$.
There is an issue about a consistent definition of
$\nu_{1,2}$ at all scales. Solar neutrino experiments
tell us that $|m_2^\lambda| > |m_1^\lambda|$, where
$\lambda$ is the laboratory scale. We {\it define}
$\nu_1$ and $\nu_2$ at higher scales in a way such that
$|m_2| \geq |m_1|$ at {\it all} scales, and in particular,
$|m_2^\Lambda| > |m_1^\Lambda|$.

In this section, we introduce four different symmetries 
and the corresponding predictions on the neutrino mixing angles 
at this high scale:

\subsection{QLC1}
\label{qlc1}

Quark-lepton complementarity 
\cite{qlc0,raidal,minakata,qlc+,ferrandis} 
links the difference between the measured
and the maximal (i.e. $45^\circ$) values of the neutrino mixing angle 
$\theta_{12}$ to the Cabbibo angle $\theta_c = 12.6^\circ \pm 
0.1^\circ$ \cite{pdg}.
We first follow a particular basis independent formulation 
``QLC1'' \cite{minakata} 
of this principle \footnote{There could be a more general 
statement of QLC1 with an additional diagonal phase matrix 
$\Gamma_\delta$ between $V_{\rm CKM}^\dagger$ and $U_{\nu, {\rm bm}}$.
But for consistency and simplicity, we choose eq.~(\ref{vdagv}).}:
\beq
U_{\rm PMNS} = V_{\rm CKM}^\dagger 
U_{\nu, {\rm bm}} \; ,
\label{vdagv}
\eeq
where $U_{\nu, {\rm bm}}$ is the specific bimaximal form \cite{bimax}
for the unitary neutrino mixing matrix.
Eq.~(\ref{vdagv}) gives rise to the ``QLC1'' relation
\footnote{We do not distinguish between $\theta_c^\Lambda$
and $\theta_c^\lambda$ since the running of $\theta_c$ is
negligible on account of the hierarchical nature of
quarks belonging to different generations.}
\beq
\theta_{12}^{\Lambda} + \frac{\theta_c}{\sqrt{2}} = 
\frac{\pi}{4} + {\cal O}(\theta_c^3) \; .
\label{qlc-form}
\eeq
The identification of (\ref{vdagv}) as a statement of QLC
becomes more transparent in the basis with $U_u = I$, i.e. where
the matrix $Y_u^\dagger Y_u$ is diagonal. 
It follows from (\ref{vckm-def}) that 
$V_{CKM} = U_d$ in this basis. 
Now a comparison of (\ref{vckm-def}) and (\ref{vdagv}),
together with the assumption of $U_\nu$ being $U_{\nu, {\rm bm}}$, 
yields the SU(5) GUT-inspired 
quark-lepton symmetry relation $U_d = U_l$.  Eq.~(\ref{vdagv}), as it
stands, is basis independent, however.

Eq. (\ref{vdagv}) yields the neutrino mixing angles at the high scale 
$\Lambda$ to be
\barr
\theta_{12}^\Lambda & = & \frac{\pi}{4} - \frac{\theta_c}{\sqrt{2}} +
{\cal O}(\theta_c^3) \approx 35.4^\circ \nonumber \\
\theta_{23}^\Lambda & = & \frac{\pi}{4} 
- |V_{cb}| -\frac{\theta_c^2}{4}  + {\cal O}(\theta_c^3)
\approx 42.1^\circ  \; , \nonumber \\
\theta_{13}^\Lambda & = &  \frac{\theta_c}{\sqrt{2}} 
+ {\cal O}(\theta_c^3) \approx 8.9^\circ \;.
\label{thetas-qlc1-high}
\earr
Thus, QLC1 predicts a value of $\theta_{13}^\Lambda$ that is 
close to the current experimental bound.

\subsection{QLC2}
\label{qlc2}

In a second version of quark-lepton complementarity, 
``QLC2'' \cite{raidal,minakata} ,
one assumes a bimaximal structure for the charged lepton mixing matrix,
$U_\ell = U_{\ell, \rm bm}$ and the form
\beq
U_{\rm PMNS} = U_{\ell, \rm bm} V_{\rm CKM}^\dagger \; ,
\label{qlc2-vdagv}
\eeq
for the PMNS matrix.
Eq.~(\ref{qlc2-vdagv}) yields in a straightforward way
the relation
\beq
\theta_{12}^{\Lambda} + \theta_c = \frac{\pi}{4} 
+ {\cal O}(\theta_c^3) \; .
\label{qlc2-t12}
\eeq
One may note that, in the basis with $U_d=I$, i.e. where 
$Y_d^\dagger Y_d$ is diagonal,
(\ref{qlc2-vdagv}) yields the SO(10) GUT-inspired relation
$U_u = U_\nu$.

Eq.~(\ref{qlc2-vdagv}) leads to the following values for the
nautrino mixing angles at the high scale:
\barr
\theta_{12}^\Lambda  & = & 
\frac{\pi}{4} - \theta_c + {\cal O}(\theta_c^3)
\approx 32.4^\circ  \; , \nonumber \\ 
\theta_{23}^\Lambda & = & \frac{\pi}{4} 
- \frac{|V_{cb}|}{\sqrt{2}} + {\cal O}(\theta_c^3)
\approx 43.4^\circ \; , \nonumber \\
\theta_{13}^\Lambda & = & \frac{|V_{cb}|}{\sqrt{2}} 
+ {\cal O}(\theta_c^3) \approx 1.6^\circ \; .
\label{thetas-qlc2-high}
\earr
The value of $\theta_{13}^\Lambda$ predicted in QLC2 is 
beyond the measuring capacity of the neutrino experiments 
planned during the next decade.

\subsection{TBM1}
\label{sec:tbm}

The tribimaximal form of 
the neutrino mixing matrix is given by 
\beq
U_{\nu, {\rm tbm}}^\Lambda = \frac{1}{\sqrt{6}} 
\left( \begin{array}{ccc}
2 & \sqrt{2} & 0 \\
-1 & \sqrt{2} & \sqrt{3} \\
1 & -\sqrt{2} & \sqrt{3} \\
\end{array} \right)
= R_{23}(\frac{\pi}{4}) R_{13}(0) R_{12}
(\sin^{-1}\frac{1}{\sqrt{3}}) \; .
\label{u-tbm}
\eeq

In the standard TBM scenario \cite{tbm}, which we refer to as TBM1,
one has $U_{\rm PMNS}^\Lambda = U_\nu^\Lambda$ 
since the charged lepton mass matrix at the high scale
is already flavour diagonal.
Then we have
\beq
\theta_{12}^\Lambda \approx 35.3^\circ
~,~
\theta_{23}^\Lambda  = 45^\circ
~,~
\theta_{13}^\Lambda  =0 ^\circ \; .
\label{thetas-tbm1-high}
\eeq

\subsection{TBM2}
\label{tbm2}

Small deviations from the tribimaximal scenario TBM1 above
have been considered in the literature \cite{tbm2,florian}, 
where the deviation
originates from the mixing in the charged lepton sector.
Here we consider the version in Ref.~\cite{tbm2}, 
and call it TBM2. Here
$U_{\rm PMNS} = V_{\rm \ell L}^\dagger U_{\nu, \rm tbm}$,
where $V_{\ell L}$ has the form \cite{ferrandis}
\beq
V_{\ell L} = \left( \begin{array}{ccc}
1 & \theta_c/3 & 0 \\
\theta_c/3 & 1 & -|V_{cb}| \\
0 & |V_{cb}| & 1 \\
\end{array} \right)
+ {\cal O}(\theta_c^3) \; ,
\label{ferrandis-form}
\eeq
with the factor of 1/3 coming from the Georgi-Jarlskog
relation \cite{GJ} $m_\mu/m_s = 3$ at the GUT scale.
As a result, we have at the high scale
\barr
\theta_{12}^\Lambda & = &
\sin^{-1} \frac{1}{\sqrt{3}} - \frac{\theta_c}{3\sqrt{2}} 
+ {\cal O}(\theta_c^3) \approx 32.3^\circ \; , \nonumber \\
\theta_{23}^\Lambda & = & 
\frac{\pi}{4} - |V_{cb}|  
+ {\cal O}(\theta_c^3) \approx 42.7^\circ \; , \nonumber \\
\theta_{13}^\Lambda  & = & \frac{\theta_c}{3\sqrt{2}} 
+ {\cal O}(\theta_c^3) \approx 3.1^\circ \; .
\label{thetas-tbm2-high}
\earr

\section{High scale structure and downward evolution}
\label{evol}

As explained in the Introduction, our idea is to start with 
a specific structure of the neutrino mass matrix 
${\cal M}_\nu^\Lambda$ that is dictated by some symmetry
at a high scale $\Lambda \sim 10^{12}$ GeV.
We would then like to evolve the elements of ${\cal M}_\nu$
down to a laboratory energy scale $\lambda \sim 10^3$ GeV.
This involves studying the (one-loop) RG evolution of the 
coefficient functions $c_{\alpha \beta}$ in
(\ref{operator}) between 
$\Lambda \sim 10^{12}$ GeV and $\lambda \sim 10^3$ GeV.
In case the considered high scale neutrino symmetries
are consequences of grand unification, we need to assume
that the threshold effects \cite{threshold} between the
GUT scale $\sim 2 \times 10^{16}$ GeV and $\Lambda$ are
flavor blind so that they do not spoil the assumed symmetry relations
in the downward evolution from $M_{\rm GUT}$ to $\Lambda$.
We also note that effects of
evolution on the masses and mixing angles 
of charged fermions are known \cite{rg-charged} to be negligibly
small \footnote{The value of $|V_{cb}|$ does run by about 0.01
in the MSSM due to the top quark U(1) coupling. However,
at the level of accuracy that we are concerned with,
this is inconsequential.} 
on account of the hierarchical nature of their mass
values.

At one loop, the neutrino mass matrices at the scales $\Lambda$
and $\lambda$ are homogeneously related \cite{chankowski,ellis-lola}:
\beq
{\cal M}_\nu^\lambda = I_K {\cal I}_\kappa^T ~ {\cal M}_\nu^\Lambda 
~{\cal I}_\kappa \; ,
\label{m-lambda}
\eeq
where 
\beq
I_K \equiv  \exp \left[ -\int_{t(\Lambda)}^{t(\lambda)} K(t) 
dt \right] \; 
\label{IK-def}
\eeq
is a scalar factor common to all elements of ${\cal M}_\nu^\lambda$.
In eq.~(\ref{IK-def}), 
$t(Q) \equiv (16 \pi^2)^{-1}  \ln(Q/Q_0)$
with $Q$ $(Q_0)$ being a running (fixed) scale,
and the integrand is given by
\beq
K(t) = -6 g_2^2(t) -2 g_Y^2(t) + 6 {\rm Tr}~(Y_u^\dagger Y_u)(t)
\eeq
in a transparent notation, $g_{2,Y}$ being the $SU(2)_L, U(1)_Y$
gauge coupling strength and $Y_u$ the up-type Yukawa coupling matrix.
Finally, the matrix $I_\kappa$ has the form
\beq
{\cal I}_\kappa \equiv \exp \left[ -  
\int_{t(\Lambda)}^{t(\lambda)} (Y_l^\dagger Y_l)(t) dt \right] \; ,
\eeq
$Y_\ell$ being the  Yukawa coupling matrix for charged leptons.

Although some of the neutrino mixing matrices in various scenarios 
in Sec.~\ref{highscale} have been motivated 
in terms of grand unification
in  bases where the symmetries involved may be clearly
observed, for the RG evolution of all scenarios
we choose to work in the basis where the charged lepton 
mass matrix is diagonal. In this basis,
\beq
Y_\ell^\dagger Y_\ell = {\rm Diag}~
(y_e^2, y_\mu^2, y_\tau^2) \; .
\label{y-ell}
\eeq
We can neglect $y_{e,\mu}^2$ in comparison with $y_\tau^2$
in (\ref{y-ell}) to get the result
\beq
{\cal I}_\kappa \approx 
\mbox{ Diag} (1,1,e^{-\Delta_\tau}) =
{\rm Diag}(1,1,1-\Delta_\tau) + {\cal O}(\Delta_\tau^2) \; ,
\label{I-kappa}
\eeq 
where \cite{chankowski}
\beq
\Delta_\tau = 
\int_{t(\Lambda)}^{t(\lambda)} |y_\tau(t) |^2 =
m_\tau^2 (\tan^2 \beta+1) (8 \pi^2 v^2)^{-1} 
\ln (\Lambda/\lambda) \; .
\label{delta-tau}
\eeq
Here 
$v \equiv \sqrt{v_u^2 + v_d^2}$
and $\tan \beta = v_u/v_d$, where $v_u$ ($v_d$) is 
$\sqrt{2}$ times the vev of 
the up (down) type neutral Higgs scalars.
Evidently, $\Delta_\tau$ is a small number for the allowed
range of $\tan\beta$: e.g. $\Delta_\tau \approx 6 \times 10^{-3}$
for $\tan\beta = 30$, justifying our neglect of the
${\cal O}(\Delta_\tau^2)$ terms \footnote{Note that a mistake of 
a factor of 2 in eq. 9 of \cite{rgqlc} has been corrected here.}.

The substitution of (\ref{I-kappa}) into (\ref{m-lambda})
leads us to 
\beq
{\cal M}_\nu^\lambda \propto \left( \begin{array}{ccc}
1 & 0 & 0 \\
0 & 1 & 0 \\
0 & 0 & 1 - \Delta_\tau \\
\end{array} \right)
{\cal M}_\nu^\Lambda
\left( \begin{array}{ccc}
1 & 0 & 0 \\
0 & 1 & 0 \\
0 & 0 & 1 - \Delta_\tau \\
\end{array} \right)
+ {\cal O}(\Delta_\tau^2) \; ,
\label{i-m-i}
\eeq
where the proportionality is through the scalar factor $I_K$
given in (\ref{IK-def}).
It must be emphasised that (\ref{i-m-i}) is valid only
when ${\cal M}_\nu^{\Lambda \, (\lambda)}$ is written in the basis where the 
charged lepton mass matrix 
is diagonal.

The matrix ${\cal M}_\nu^\Lambda$ is complex symmetric.
Writing it in the general form
\beq
{\cal M}_\nu^\Lambda = \left( \begin{array}{ccc}
A & B & C \\
B & D & E \\
C & E & F \\
\end{array} \right) \; ,
\label{abc-form-highscale}
\eeq
we have
\beq
{\cal M}_\nu^\lambda = \left( \begin{array}{ccc}
A & B & C (1 - \Delta_\tau)\\
B & D & E (1 - \Delta_\tau)\\
C (1 - \Delta_\tau)& E (1 - \Delta_\tau)& F (1 - 2\Delta_\tau)\\
\end{array} \right) 
+ {\cal O}(\Delta_\tau^2) \; .
\label{abc-form-lowscale}
\eeq
Since both $M_\nu^{\Lambda, \lambda}$ are complex 
symmetric matrices, they can be diagonalized as
\bea
(U^\Lambda)^\dagger ~M_\nu^\Lambda (U^\Lambda)^*
& = & {\rm Diag}(|m_1^\Lambda|, |m_2^\Lambda|, |m_3^\Lambda|) \; ,
\nonumber \\ 
(U^\lambda)^\dagger ~M_\nu^\lambda (U^\lambda)^*
& = & {\rm Diag}(|m_1^\lambda|, |m_2^\lambda|, |m_3^\lambda|) \; .
\eea
Note that, since we are working in a basis where the charged
lepton mass matrix is diagonal, $U^{\Lambda \; (\lambda)}$ is the 
same as the net leptonic mixing matrix $U_{\rm PMNS}$
(\ref{upmns-def}) at the scale $\Lambda \; (\lambda)$.

The unitary matrix $U_{\rm PMNS}$ may be parametrised in its
most general form as
\begin{eqnarray}
U_{\rm PMNS} & \equiv & 
{\rm Diag}(e^{i\phi_e}, e^{i\phi_\mu}, e^{i\phi_\tau})
R_{23}(\theta_{23})
{\rm Diag}(1, 1, e^{i \delta_\ell}) 
R_{13}(\theta_{13}) \times \nonumber \\
& & {\rm Diag}(1, 1, e^{-i \delta_\ell}) R_{12}(\theta_{12})
{\rm Diag}(1, e^{-i\alpha_2/2}, e^{-i\alpha_3/2}) \, , 
\label{u-lambda}
\end{eqnarray}
where $R_{ij}(\theta_{ij})$ is the matrix for rotation 
through the angle $\theta_{ij}$ in the $i-j$ plane,
$\delta_\ell$ is the CP violating Dirac phase, 
$\alpha_{2,3}$ the Majorana phases and $\phi_{e,\mu,\tau}$
the so-called ``unphysical'' additional phases required
to diagonalise the neutrino mass matrix.
Here we have already used the freedom of choosing
$\alpha_1=0$.

If $\theta_{13}^\Lambda$ vanishes on account of the 
symmetry requirement at the scale $\Lambda$,
the evolution of the mixing angles can be computed 
analytically in a simple manner.
The matrix $U^\Lambda$ can then be written as 
\beq
U^\Lambda = R_{23}(\theta_{23}^\Lambda)
R_{12}(\theta_{12}^\Lambda)
{\rm Diag}(1, e^{-i\alpha_2^\Lambda/2},e^{-i\alpha_3^\Lambda/2})\; ,
\label{u-high}
\eeq
since the  Dirac phase contribution vanishes and the phases
$\phi_{e,\mu,\tau}$ can anyway be absorbed in the 
charged lepton phases.
RG evolution will modify the angles $\theta_{23}, \theta_{12}$
as well as the phases $\alpha_{2,3}$, 
at the same time generating nonzero values for the mixing angle
$\theta_{13}$ and the Dirac phase $\delta_\ell$. The phases
$\phi_{e,\mu,\tau}$ that may get generated can always be absorbed
in the phases of the charged lepton flavour eigenstates.

If we now approximate the deviation of $U_\nu^\lambda$
from $U_\nu^\Lambda$ by the retention of only terms that are
linear in $\Delta_\tau$, we can write the modified mixing angles
as
\beq
\theta_{12}^{\lambda} = \theta_{12}^{\Lambda} + 
k_{12} \Delta_\tau 
+ {\cal O}(\Delta_\tau^2) ~ , ~ 
\theta_{23}^{\lambda} = \theta_{23}^{\Lambda} + 
k_{23} \Delta_\tau
+ {\cal O}(\Delta_\tau^2) ~ ,~
\theta_{13}^{\lambda}  =  
k_{13} \Delta_\tau + {\cal O}(\Delta_\tau^2) \;.
\label{thetaijs}
\eeq
The modified phases may similarly be written as
\beq
\alpha_{2,3}^\lambda = \alpha_{2,3}^\Lambda 
+ a_{2,3} \Delta_\tau + {\cal O}(\Delta_\tau^2) ~,~
\delta_\ell^\lambda = d_\ell 
\Delta_\tau + {\cal O}(\Delta_\tau^2) \; ,
\label{alpha-evol}
\eeq
where we expect $a_{2,3}$ and $d_\ell$ to be
${\cal O}(1)$ quantities.
In this paper we shall only be concerned about the
deviation $k_{ij}\Delta_\tau$ of the mixing angles
$\theta_{ij}$ from their high scale values.
We shall see a posteriori that $|k_{ij} \Delta_\tau|$ is always 
much less than unity so that we can ignore its quadratic and higher
powers.

The values of $k_{ij}$ are found to be
\barr
k_{12} & = & \frac{1}{2} 
\sin 2\theta_{12}^{\Lambda} \sin^2 \theta_{23}^{\Lambda}
\frac{|m_1^\Lambda + m_2^\Lambda|^2}{|m_2^\Lambda|^2 - |m_1^\Lambda|^2} 
\; , \nonumber \\
k_{23} & = & \frac{1}{2} \sin 2 \theta_{23}^{\Lambda} \left( 
\cos^2 \theta_{12}^{\Lambda} \frac{|m_2^\Lambda + m_3^\Lambda|^2}{
|m_3^\Lambda|^2 - |m_2^\Lambda|^2} +
\sin^2 \theta_{12}^{\Lambda} \frac{|m_1^\Lambda + m_3^\Lambda|^2}{
|m_3^\Lambda|^2 - |m_1^\Lambda|^2} 
\right) \;, \nonumber \\
k_{13} & = & \frac{1}{4} 
\sin 2\theta_{12}^{\Lambda} \sin 2 \theta_{23}^{\Lambda} \left( 
\frac{|m_2^\Lambda + m_3^\Lambda|^2}{|m_3^\Lambda|^2 - |m_2^\Lambda|^2} -
\frac{|m_1^\Lambda + m_3^\Lambda|^2}{|m_3^\Lambda|^2 - |m_1^\Lambda|^2} 
\right) \; ,
\phantom{space}
\label{kij-m1m2m3}
\earr
where $m_i^\Lambda$ are the masses $|m_i^\Lambda| e^{i\alpha_i^\Lambda}$
at the high scale. 
The RG evolution of $|m_i|$ may  be
parametrised by
\beq
|m_i^\lambda| = I_K |m_i^\Lambda| \left( 1 + \mu_i \Delta_\tau + {\cal O}
(\Delta_\tau^2)\right)
\label{m-evol}
\eeq
where $I_K$ is the scalar factor given in (\ref{IK-def})
and $\mu_i$ are ${\cal O}(1)$ numbers \cite{antusch-majorana}.
Then, taking $m_i$ to be the masses $|m_i^\lambda|
e^{i\alpha_i^\lambda}$ at the low scale introduces
an error of ${\cal O}(\Delta_\tau)$ in $k_{ij}$,
and hence of ${\cal O}(\Delta_\tau^2)$ in $\theta_{ij}$.
Eqs.~(\ref{kij-m1m2m3}) are therefore valid even with
the low scale values of the masses and Majorana phases.
The same argument is true even for the mixing angles
$\theta_{ij}$. 
Since all observations in the neutrino experiments
are made at laboratory energies, we henceforth drop the
superscript $\lambda$ for all the quantities at the low
scale.

The expressions (\ref{kij-m1m2m3}) have been derived starting
with $\theta_{13}^\Lambda =0$, and they agree with the 
general expressions in \cite{antusch-majorana} 
in the limit $\theta_{13}^\Lambda \to 0$.
For finite $\theta_{13}^\Lambda$, the $k_{ij}$ given in
(\ref{kij-m1m2m3}) have an error of 
${\cal O}(\theta_{13}^\Lambda)$, i.e. the mixing
angles at the low scale have an error of 
${\cal O}(\theta_{13}^\Lambda \Delta_\tau)$,
which we neglect in our analytical approximations.
Since for all the scenarios considered here,
$\theta_{13}^\Lambda \Delta_\tau \lsim 0.1^\circ$,
this is a justified assumption.
Thus, even with small nonzero values of $\theta_{13}^\Lambda$,
as is the case with QLC1, QLC2 and TBM2 scenarios,
we can justifiably use the result
$\theta_{ij} \approx \theta_{ij}^\Lambda + k_{ij} \Delta_\tau$
with $k_{ij}$ given by (\ref{kij-m1m2m3}).
Another  requirement for the validity of 
eqs. (\ref{kij-m1m2m3}) is that the values of
$m_i$ and $|m_i|^2-|m_j|^2$ should be described accurately by
the ${\cal O}(\Delta_\tau)$ terms in their RG evolution.
This condition does not impose any restriction on $m_i^\Lambda$.
However, for
$\Delta_\tau \gsim (|m_2^\Lambda|^2-|m_1^\Lambda|^2)/(m_0^\Lambda)^2$, 
the ${\cal O}(\Delta_\tau^2)$ terms dominate over the ${\cal O}
(\Delta_\tau)$ terms in the $|m_2|^2-|m_1|^2$ evolution
\cite{antusch-majorana}, which results in
eqs.~(\ref{kij-m1m2m3}) breaking down.
Thus for the validity of these equations, we require
\beq
(m_0^\Lambda)^2  \cdot \Delta_\tau \lsim |m_2^\Lambda|^2 - |m_1^\Lambda|^2 \;.
\label{validity}
\eeq
Eq.~(\ref{validity}) may be violated if 
$|m_2^\Lambda|^2 - |m_1^\Lambda|^2$ is indeed very small. 
However, even in such a situation, many of the qualitative
features following from our analytical treatment will
continue to remain valid. This will be confirmed by the
exact numerical analysis given later.

One can make some further observations on eqs.~(\ref{kij-m1m2m3}).
Since $k_{12}$ is always positive, the measured value of 
$\theta_{12}$ can never be smaller than $\theta_{12}^\Lambda$.
Moreover, in order to have any significant effect of 
RG evolution, some $|k_{ij}|$
are required to be $\gsim {\cal O}(10)$ since 
$\Delta_\tau \lsim {\cal O}(10^{-2})$. 
This requirement is not met in the case of
a normal hierarchy which has all $|k_{ij}|\sim {\cal O}(1)$,
and hence there is no significant RG effect there.
Specifically, for an enhancement of $|k_{12}|$, $|m_0^\Lambda|^2$ must
necessarily be $\gg |m_2^\Lambda|^2 - |m_1^\Lambda|^2$.
This inequality is obeyed whenever neutrinos are
nonhierarchical, i.e. either in the IH or the QD case.
On the other hand, the enhancement of both $|k_{23}|$ and
$|k_{13}|$ requires the inequality 
$|\rho_A^\Lambda| \ll 1$,
satisfied only by QD neutrinos.
However, these are necessary conditions. The occurence (or not)
of an actual enhancement depends on the Majorana phases
$\alpha_{2,3}$ too.
In order to see this dependence explicitly,
let us rewrite (\ref{kij-m1m2m3}) in terms of 
these phases and our parameters $\rho_A^\Lambda$ and 
$\epsilon_S^\Lambda$ of 
(\ref{rhoeps-def}), defined at the high scale.
We then have, with finite $\theta_{13}^\Lambda$,
\barr
k_{12} & = & \frac{1}{4 \epsilon_S^\Lambda} 
\sin 2 \theta_{12}^{\Lambda} \sin^2 \theta_{23}^{\Lambda} \left[
1+\cos\bb^\Lambda + (\epsilon_S^\Lambda)^2(1-\cos\bb^\Lambda) \right]  
+ {\cal O}(\theta_{13}^\Lambda) 
\; , \nonumber \\
k_{23} & = & \frac{\Gamma^\Lambda}{4} 
\sin 2\theta_{23}^{\Lambda} 
\left[ 1 + \cos^2 \theta_{12}^{\Lambda} \cos(\bb^\Lambda-\cc^\Lambda) + 
\sin^2 \theta_{12}^{\Lambda} \cos\cc^\Lambda \right]  
\nonumber \\
& & \phantom{theta} + \frac{\rho_A^\Lambda}{2}
\sin 2\theta_{12}^{\Lambda} 
\sin 2\theta_{23}^{\Lambda} 
+ {\cal O}(\epsilon_S^\Lambda, \theta_{13}^\Lambda) \; , 
\phantom{spa} \nonumber \\
k_{13} & = & \frac{\Gamma^\Lambda}{8} 
\sin 2\theta_{12}^{\Lambda} \sin 2 \theta_{23}^{\Lambda} \left[
\cos(\bb^\Lambda-\cc^\Lambda) 
- \cos\cc^\Lambda \right] + {\cal O}(\epsilon_S^\Lambda, 
\theta_{13}^\Lambda) \; .
\label{kij-rhoeps}
\earr
It is clear from (\ref{kij-rhoeps}) that
the values of the Majorana phases $\alpha_{2,3}^\Lambda$ are
crucial in controlling whether the evolution terms 
$k_{ij} \Delta_\tau$ are dangerously large or not.
For instance, as $\alpha_2^\Lambda \to 0$, we have $k_{12}
\approx (4 \epsilon_S^\Lambda)^{-1}$.
For nonhierarchical neutrinos, the latter is likely to destroy any 
high scale symmetry statement on $\theta_{12}$.
When $\alpha_2^\Lambda$ increases from zero, $|k_{12}|$
decreases rapidly, becoming as small as $|\epsilon_S^\Lambda|/4$
when $\alpha_2^\Lambda =\pi$, i.e. $m_1^\Lambda \approx -m_2^\Lambda$.
Moreover, (\ref{kij-rhoeps}) also indicates that 
both $|k_{23}|$ and $|k_{13}|$
are enhanced when either $\alpha_3^\Lambda=0$, 
or $\alpha_2^\Lambda = \alpha_3^\Lambda$
for a nonzero $\alpha_2^\Lambda$, though $|k_{13}|$
gets highly suppressed when $\alpha_2^\Lambda =0$.

As noted in Sec.~\ref{param}, 
the sign of $\Gamma^\Lambda$ is positive for a normal 
ordering and negative  for an inverted ordering of the
neutrino masses.
Thus $k_{23}$ is always positive (negative) for a 
normal (inverted) ordering. 
However, the sign of $k_{13}$ is controlled not only by
${\rm sgn}(\Gamma^\Lambda)$ but also by a combination of 
Majorana phases.

In the remaining part of this section, we enumerate the predictions
for the four scenarios considered in this paper.
On evolution down to the scale $\lambda$, 
the mixing angles of the corresponding mixing matrix 
$U_{\rm PMNS}$
are given in all the four cases by 
$\theta_{ij} = \theta_{ij}^{\Lambda} 
+ k_{ij} \Delta_\tau + 
{\cal O}(\theta_{13}^\Lambda \Delta_\tau, \Delta_\tau^2)$
where 
the values of $k_{ij}$ are given by (\ref{kij-rhoeps}).

\subsection{QLC1}
\label{qlc1-evolved}

On evolving (\ref{thetas-qlc1-high}) to the laboratory scale, 
the net leptonic mixing angles are found to be
\barr
\theta_{12}  & = & \frac{\pi}{4} -
\frac{\theta_c}{\sqrt{2}} 
+ k_{12}^{\rm QLC1} \Delta_\tau 
+ {\cal O}(\theta_c^3,  
\theta_{13}^\Lambda \Delta_\tau, \Delta_\tau^2) \; , \nonumber \\ 
\theta_{23} & = & \frac{\pi}{4} 
- |V_{cb}| -\frac{\theta_c^2}{4} + k_{23}^{\rm QLC1} \Delta_\tau
+ {\cal O}(\theta_c^3, 
\theta_{13}^\Lambda \Delta_\tau, \Delta_\tau^2) \; , \nonumber \\ 
\theta_{13} & = & \left| \frac{\theta_c}{\sqrt{2}} 
- k_{13}^{\rm QLC1} \Delta_\tau \right|
+ {\cal O}(\theta_c^3, 
\theta_{13}^\Lambda \Delta_\tau, \Delta_\tau^2) \; ,
\label{thetas-qlc1}
\earr
where the neglected terms are $\lsim 0.1^\circ$.
The absolute value taken for the RHS of $\theta_{13}$ is in order to
keep to the convention of defining $\theta_{ij} >0$ \cite{pdg}.
 
Since 
$\pi/4 - \theta_c/\sqrt{2} \approx 35.4^\circ$ and
$k_{12}^{\rm QLC1} \Delta_\tau > 0$, it follows from 
(\ref{thetas-qlc1}) that $\theta_{12}$ in this scenario
cannot be less than $35.4^\circ$.
Further, since 
$\pi/4 - |V_{cb}| - \theta_c^2/4 \approx 42.1^\circ$, and 
$k_{23}^{\rm QLC1} \Delta_\tau >0$ ($<0$) for a normal
(inverted) ordering of neutrino masses, a consequence of
(\ref{thetas-qlc1}) is that $\theta_{23}$ is greater than 
(less than) $42.1^\circ$ for a normal (inverted) ordering. 
Finally, the predicted value of $\theta_{13}$ is 
$\theta_c/\sqrt{2} \approx 8.9^\circ$
in the absence of RG running, but it can be greater or less than
$8.9^\circ$ depending on the values of the Majorana phases.
The detailed numerical analysis will be presented in 
Sec.~\ref{constraints}.

\subsection{QLC2}
\label{qlc2-evolved}

After RG evolution, the high scale angles (\ref{thetas-qlc2-high}) 
in this scenario evolve to
\barr
\theta_{12}  & = & \frac{\pi}{4} -\theta_c
+ k_{12}^{\rm QLC2} \Delta_\tau -  
+ {\cal O}(\theta_c^3, 
\theta_{13}^\Lambda \Delta_\tau, \Delta_\tau^2) \; , \nonumber \\
\theta_{23} & = & \frac{\pi}{4} 
- \frac{|V_{cb}|}{\sqrt{2}}
+ k_{23}^{\rm QLC2} \Delta_\tau
+ {\cal O}(\theta_c^3, 
\theta_{13}^\Lambda \Delta_\tau, \Delta_\tau^2) \; , \nonumber \\
\theta_{13} & = & \left|\frac{|V_{cb}|}{\sqrt{2}} 
- k_{13}^{\rm QLC2} \Delta_\tau \right|
+ {\cal O}(\theta_c^3, 
\theta_{13}^\Lambda \Delta_\tau, \Delta_\tau^2) \; ,
\label{thetas-qlc2}
\earr
where the neglected terms are $\lsim 0.1^\circ$ 
as before.

The lower bound on $\theta_{12}$ in this scanario being
$\pi/4 - \theta_c \approx 32.4^\circ$, significantly lower
values of $\theta_{12}$ than QLC1 are allowed.
Also, here $\theta_{23}$ is greater than (less than) 
$\pi/4 - |V_{cb}|/\sqrt{2} \approx 43.4^\circ$
for normal (inverted) neutrino mass ordering. 
The major difference from QLC1 is
in $\theta_{13}$: the value $\theta_{13}$ in QLC2 is only
$|V_{cb}|/\sqrt{2} \approx 1.6^\circ$ in the absence of
RG running. It can increase with RG running, but the extent
of this increase is restricted from the observed $\theta_{12}$
values which restrict the values of the Majorana phases in turn.
The detailed numerical analysis
will again be presented in Sec.~\ref{constraints}.

\subsection{TBM1}
\label{tbm1-evolved}

The mixing angles at the low scale here are simply given by
\barr
\theta_{12} & = &
\sin^{-1} \frac{1}{\sqrt{3}} 
+ k_{12}^{\rm TBM1} \Delta_\tau
+ {\cal O}(\theta_c^3, 
\theta_{13}^\Lambda \Delta_\tau, \Delta_\tau^2) \; , \nonumber \\
\theta_{23} & = & 
\frac{\pi}{4} 
+ k_{23}^{\rm TBM1} \Delta_\tau 
+ {\cal O}(\theta_c^3, 
\theta_{13}^\Lambda \Delta_\tau, \Delta_\tau^2) \; , \nonumber \\
\theta_{13}  & = & k_{13}^{\rm TBM1} \Delta_\tau 
+ {\cal O}(\theta_c^3, 
\theta_{13}^\Lambda \Delta_\tau, \Delta_\tau^2) \; .
\label{thetas-tbm1}
\earr
On similar lines to the arguments given for the QLC scenarios,
here (i) the minimum value of $\theta_{12}$ is 
$\sin^{-1}(1/\sqrt{3}) \approx 35.3^\circ$, 
(ii) the value of $\theta_{23}$ is greater than
(less than) $45^\circ$ for normal (inverted) hierarchy, and the
value of $\theta_{13}$ vanishes in the absence of RG running.
Since (\ref{kij-rhoeps}) shows that $\theta_{13}$ does not run
if the Majorana phases vanish, {\it any observed deviation
of $\theta_{13}$ from zero in this scheme will indicate nonvanishing
Majorana phases.}
The detailed numerical analysis appears in Sec.~\ref{constraints}.

\subsection{TBM2}
\label{tbm2-evolved}

The mixing angles 
(\ref{thetas-tbm2-high}) for this scenario evolve 
to the laboratory scale and become
\barr
\theta_{12} & = &
\sin^{-1} \frac{1}{\sqrt{3}} 
- \frac{\theta_c}{3\sqrt{2}} 
+ k_{12}^{\rm TBM2} \Delta_\tau
+ {\cal O}(\theta_c^3, 
\theta_{13}^\Lambda \Delta_\tau, \Delta_\tau^2) \; , \nonumber \\
\theta_{23} & = & 
\frac{\pi}{4} - |V_{cb}|
+ k_{23}^{\rm TBM2} \Delta_\tau 
+ {\cal O}(\theta_c^3, 
\theta_{13}^\Lambda \Delta_\tau, \Delta_\tau^2) \; , \nonumber \\
\theta_{13}  & = & \left| \frac{\theta_c}{3\sqrt{2}}
-k_{13}^{\rm TBM2} \Delta_\tau \right|  
+ {\cal O}(\theta_c^3, 
\theta_{13}^\Lambda \Delta_\tau, \Delta_\tau^2) \; .
\label{thetas-tbm2}
\earr
The minimum allowed value of $\theta_{12}$ in this
scenario is 
$\sin^{-1}(1/\sqrt{3}) - \theta_c/(3\sqrt{2})
\approx 32.3^\circ$, since $k_{12}^{\rm TBM2} >0$.
The value of $\theta_{23}$ is greater 
than (less than) 
$\pi/4 - |V_{cb}| 
\approx 42.7^\circ$
for normal (inverted) hierarchy.
Finally, the value of 
$\theta_{13}$ is 
$\theta_c/(3\sqrt{2}) 
\approx 3.1^\circ$ 
in the absence of RG running.
Again, the detailed numerical analysis with RG running is presented 
in Sec.~\ref{constraints}.

\section{Constraints on $m_0$, $\tan \beta$ and Majorana phases}
\label{constraints}

In this section, we explore the current limits on the
parameters of the four scenarios considered above. 
The main parameters governing RG running are
$m_0$, $\tan\beta$, the Majorana phases $\alpha_{2,3}$,
and (to a smaller extent) the Dirac phase $\delta_\ell$.
Our aim is to find the range of values of these parameters
allowed by the current data.
The four scenarios also lead to slightly different predictions for 
the mixing angles; and accurate measurements of these
angles should distinguish among them in the absence
of RG running. However, since the latter spoils  
high scale symmetries in general, and since the values of
all the relevant parameters are still not known, 
the low scale predictions of the mixing angles are expected
to ovelap. We explore in detail whether this still
allows one to discriminate among the scenarios
considered in this paper.
Moreover, we study the correlations between the
deviations of the mixing angles from their high scale
values.

We use the $3\sigma$ ranges for the neutrino 
mass and mixing parameters
\beq
7 \times 10^{-5} ~{\rm eV}^2  < \delta m^2 _S < 
9.1 \times 10^{-5} ~{\rm eV}^2 \; , \quad
1.7 \times 10^{-3} ~{\rm eV}^2  < |\delta m^2 _A| < 
 3.3 \times 10^{-3} ~{\rm eV}^2 \; , 
\label{low-dmsq}
\eeq
\beq
30^\circ < \theta_{12} <  39.2^\circ \; , \quad
35.5^\circ < \theta_{23} <  55.5^\circ \; , \quad
\phantom{<} \theta_{13} <  12^\circ \;  
\label{low-angles}
\eeq
at the low scale. At the high scale, we start with the values of
$\theta_{ij}^\Lambda$ dictated by the scenario under consideration,
and a range of $\delta m^2_{S/A}$ values that are consistent,
after RG evolution, 
with the low scale measurements (\ref{low-dmsq}) and 
(\ref{low-angles}).
We show our results for a normal mass ordering of neutrinos. 
However, the constraints in the case of an inverted mass 
ordering are almost identical.

We give our constraints in terms of the values of $m_0^\Lambda$
and $\alpha_2^\Lambda$ at the high scale. These are related to
the low scale values of $m_0$ and $\alpha_2$ as indicated in
(\ref{m-evol}) and (\ref{alpha-evol}) respectively. In particular,
the evolution of $m_0$ is controlled mainly by $I_K \approx 0.71$,
which makes $m_0^\lambda \approx 0.71 \, m_0^\Lambda$. 
The bounds on 
$m_0^\Lambda$ shown in the figures in this section can then be
easily translated into bounds on the low scale value of $m_0$.
The RG evolution of $\alpha_2$ is given by 
$\alpha_2^\lambda \approx \alpha_2^\Lambda + a_2 \Delta_\tau$ with
\cite{antusch-majorana}
\beq
a_2 \approx \frac{- 4 |m_1^\Lambda m_2^\Lambda|}{|m_2^\Lambda|^2 -
|m_1^\Lambda|^2} \cos 2\theta_{12}^\Lambda \sin^2 \theta_{23}^\Lambda
\sin \alpha_2^\Lambda \; ,
\label{a2-expr}
\eeq
which may be used to translate the $\alpha_2^\Lambda$ constraints
to low scale values of $\alpha_2$.

We have neglected possible Planck scale effects \cite{planck}
which may change the value of $\theta_{12}$ by a few degrees
for quasidegenerate neutrinos, leaving the other two angles
virtually unaffected. Inclusion of these effects would
relax \cite{planck-dgr} the constraints 
in the $m_0^\Lambda$ -- $\alpha_2^\Lambda$ 
plane by a small amount.

\subsection{Limits on $m_0^\Lambda$, $\alpha_2^\Lambda$ 
and $\tan \beta$}
\label{m0alpha}

Out of the three leptonic mixing angles, $\theta_{12}$ is
the one measured with the greatest accuracy currently.
Since all the scenarios have specific predictions for
$\theta_{12}$ in the absence of RG evolution, the
measured value of $\theta_{12}$ can put the strongest
constraints on the running parameters.
From (\ref{thetaijs}) and (\ref{kij-rhoeps}), 
the running of $\theta_{12}$ is expected to be 
independent of $\alpha_3^\Lambda$.

\begin{figure}[t]
\epsfig{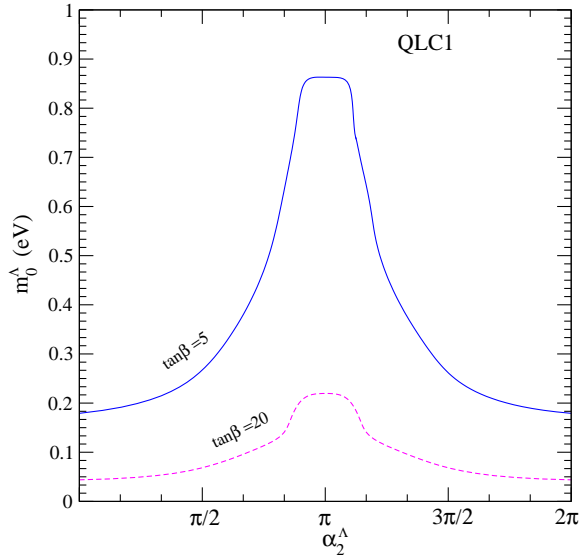}
\caption{
Constraints from the $3\sigma$ allowed range of $\theta_{12}$ 
on the $m_0$ (eV) -- $\bb$ (radians)
parameter space for  $\tan\beta$ = 5 and 20
in the QLC1 scenario.
The regions above the contours are excluded by data 
for that particular value of $\tan\beta$.
The peak at $\alpha_2 \approx \pi$ is noteworthy.
\label{mbeta-qlc1}
}
\end{figure}

We show in Fig.~\ref{mbeta-qlc1} the $3\sigma$
allowed regions in the $m_0^\Lambda-\bb^\Lambda$ plane for two
$\tan\beta$ values.
This figure has been obtained from numerical solutions 
of the RG equations.
The figure agrees very well with our expectations 
from the analytic expressions (\ref{kij-rhoeps}).
On account of the occurence of the small quantity $\epsilon_S$
in the denominator of $k_{12}$ in eq.~(\ref{kij-rhoeps}), 
strong constraints ensue on $m_0^\Lambda$ and $\alpha_2^\Lambda$.
At large $\tan \beta$ 
(equivalent to a relatively large $\Delta_\tau$), 
the value of $\bb^\Lambda$ has to be near $\pi$ 
(i.e. $m_1^\Lambda \approx -m_2^\Lambda$) 
underscoring the necessity for a nontrivial Majorana phase.
However, the requirement is less severe for a smaller 
$\tan\beta$.
The coefficient $a_2$ in (\ref{a2-expr}) characterising
the evolution of $\alpha_2$ vanishes when 
$\alpha_2^\Lambda = \pi$ \cite{antusch-majorana}. 
As a result, the preferred value of $\alpha_2$,
viz. $\alpha_2^\Lambda=\pi$, is equivalent
to $\alpha_2 =\pi$ at all scales.

\begin{figure}[t]
\epsfig{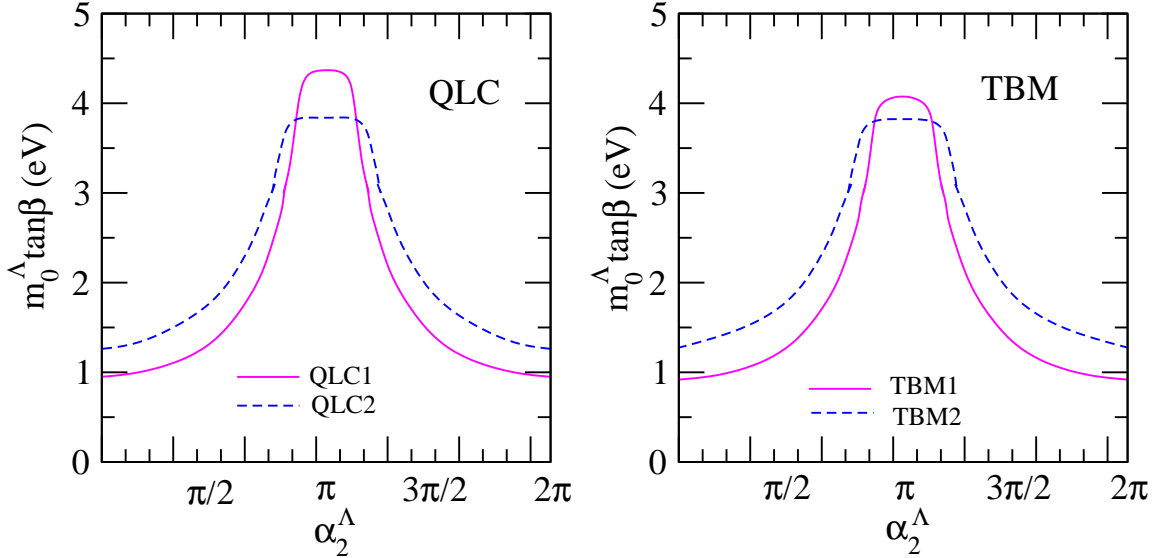}
\caption{Constraints from the $3\sigma$ allowed range of
$\theta_{12}$ on the $m_0\tan \beta$ (eV) -- $\bb$ (radians)
parameter space for the four scenarios.
The regions above the contours are excluded by data
for that particular scenario.
\label{mtanbeta-fig}
}
\end{figure}

The constraints displayed in the figure 
are calculated for two fixed $\tan\beta$ values.
The results for other values of $\tan\beta$ may be 
extrapolated from the figure.
However, we observe numerically that the $\theta_{12}$
constraints on the $m_0^\Lambda$ -- $\alpha_2^\Lambda$ plane
(Fig.~\ref{mbeta-qlc1}) depend essentially on the
combination $m_0^\Lambda \tan\beta$.
Therefore, in Fig.~\ref{mtanbeta-fig}, we show 
the allowed region in the $m_0 \tan\beta$ -- $\alpha_2$
parameter space for all four scenarios. 
The differences among the four
scenarios arise primarily from the differences in the 
values of $\theta_{12}^\Lambda$.
These are $35.4^\circ$ (QLC1), $32.4^\circ$ (QLC2),
$35.3^\circ$ (TBM1) and $32.3^\circ$ (TBM2)
in the four scenarios.
The deviation of $\theta_{12}$ from this value, 
$\Delta\theta_{12} \equiv \theta_{12} - \theta_{12}^\Lambda$ 
differs in all the scenarios only through
a factor of $\sin 2 \theta_{12}^\Lambda 
\sin^2 \theta_{23}^\Lambda$ [see 
eq.~(\ref{kij-rhoeps})], which is equal to unity within 
5\% for all the four scenarios.
As a result, the allowed regions for QLC2 and TBM2 are nearly
identical, and larger than those for QLC1 and TBM1,
the last two regions being also almost identical
to each other.

Note that, for all the scenarios, the larger the value of
$m_0^\Lambda \tan\beta$, the closer the value of $\alpha_2^\Lambda$
needs to be to $\pi$.
This statement is true even if we use the laboratory
values of $m_0$ and $\alpha_2$.
Moreover, the region with $m_0^\Lambda \tan\beta \gsim 4.4$ eV
(i.e. $m_0 \tan\beta \gsim 3.1$ eV)
is disallowed for all values of $\alpha_2$ and for
all scenarios.

\subsection{$\theta_{13}$ to discriminate among scenarios}
\label{theta13}

The four scenarios that we consider here give different 
predictions for the values of $\theta_{13}^\Lambda$:
$8.9^\circ$ (QLC1), $1.6^\circ$ (QLC2), $0^\circ$ (TBM1)
and $3.1^\circ$ (TBM2).
In the absence of RG running, therefore, a discrimination
between some of these scenarios should be possible in
the near future. For example, if QLC1 is realised in nature,
the value of $\theta_{13}$ would be accessible to the 
current generation of experiments.
However, the value of $\theta_{13}$ changes with RG
evolution and can either increase or decrease depending
on the values of Majorana phases, as can be seen from
(\ref{kij-rhoeps}).
It is conceivable that the allowed ranges of $\theta_{13}$
values for all the scenarios will then overlap and 
the power of discrimination will be lost.   
It is thus worthwhile to check whether one retains
this discrimination capability in spite of 
the RG running. What helps in this is the fact that 
at higher values of $m_0^\Lambda \tan\beta$, where one expects
large RG effects, not all $\alpha_{2,3}^\Lambda$ values are allowed:
the observed values of $\theta_{12}$ (as shown in 
Sec.~\ref{m0alpha}) as well as of $\theta_{23}$ restrict
the values of the Majorana phases, which in turn restrict
the allowed values of $\theta_{13}$.

\begin{figure}
\epsfig{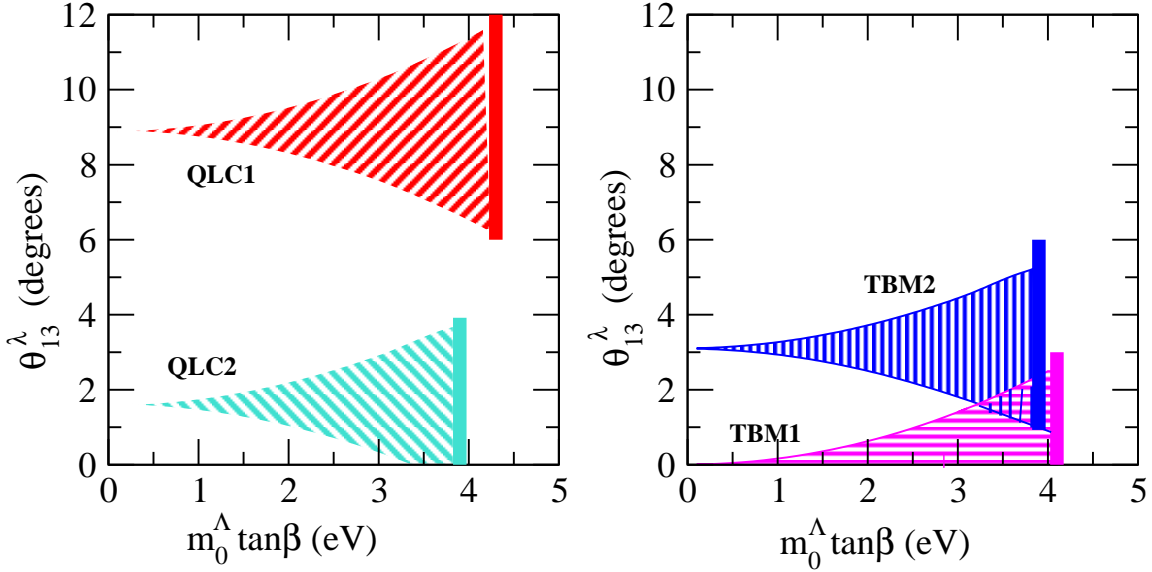}
\caption{Values of $\theta_{13}$ in the QLC scenarios
(left panel) and the TBM scenarios (right panel)
as a function of $m_0 \tan\beta$, allowed with
the current constraints on the mixing angles and
mass squared differences.
The shaded region represents the allowed values of 
$\theta_{13}$ at the laboratory scale in each scenario. 
The thick bands represent upper limits on 
$m_0^\Lambda \tan\beta$ from current measurements.
\label{theta13-fig}
}
\end{figure}

In Fig.~\ref{theta13-fig}, we show the pemitted values of 
$\theta_{13}$ in the four scenarios, subject to the constraints
of the $3\sigma$ allowed current ranges of the mixing angles.
With the current constraints, it should be possible to
distinguish between QLC1 and the other scenarios
in the next round of experiments \cite{theta13-expts}
probing $\theta_{13}$.
For example, if $\theta_{13}$ is shown to be less than
$6^\circ$, QLC1 will be excluded. 
 
The scenario TBM2 may be distinguishable from the remaining
two cases if $m_0^\Lambda \tan\beta \lsim 2$ eV 
(i.e. $m_0 \tan\beta \lsim 1.4$ eV).
So this needs, in addition to the information on $\theta_{13}$,
stronger constraints on the values of $m_0$ or $\tan\beta$.
A more accurate measurement \cite{theta12-expt}
of $\theta_{12}$ will also indirectly
limit the extent of RG evolution and may enable one to
distinguish TBM2 from the other two scenarios, viz. QLC2 and TBM1.
The last two scenarios have very similar predictions 
on $\theta_{13}$.
This is on account of the close values of 
$\theta_{13}^\Lambda$ in the two scenarios, and the 
nearly identical evolution
$\Delta\theta_{13} \equiv \theta_{13}-\theta_{13}^\Lambda$ 
in them, whatever the values of $\alpha_2^\Lambda, 
\alpha_3^\Lambda, m_0^\Lambda$ and 
$\tan\beta$.
From (\ref{kij-rhoeps}), the value of
$\Delta\theta_{13}$
differs in all the scenarios only through
a factor of $\sin 2 \theta_{12}^\Lambda 
\sin 2 \theta_{23}^\Lambda$ [see 
eq.~(\ref{kij-rhoeps})], which is equal to unity within 
5\% for all the four scenarios.
Therefore, distinguishing between these two scenarios
on the basis of a measurement of $\theta_{13}$ alone will 
be difficult unless the values of the other parameters --
$m_0^\Lambda, \tan\beta, \alpha_2^\Lambda, \alpha_3^\Lambda$ 
-- are also known to a good accuracy.

Though the constraints on $m_0^\Lambda \tan\beta$ and
$\alpha_2^\Lambda$ in sec.~\ref{m0alpha} are logarithmically 
sensitive to the choice of the high scale $\Lambda$,
the limits on $\theta_{13}$ in a given scenario are almost
indepenent of $\Lambda$. This is because the maximum
allowed value of $|\Delta\theta_{13}|$ is dictated mainly by the
maximum allowed value of $\Delta\theta_{12}$
through $|\Delta\theta_{13}|/\Delta\theta_{12}
\approx |k_{13}|/k_{12} $, whereas the maximum allowed value of
$\Delta\theta_{12}$ in turn is an experimentally determined
quantity, quite independent of $\Lambda$.

\subsection{Effect of RG evolution on $\theta_{23}$}
\label{theta23}

The predictions of all the scenarios for $\theta_{23}$ are almost
identical. 
The values of $\theta_{23}^\Lambda$ are very close:
$42.1^\circ$ (QLC1), $43.4^\circ$ (QLC2), $45^\circ$ (TBM1)
and $42.7^\circ$ (TBM2).
Moreover, the deviations
$\Delta\theta_{23} \equiv \theta_{23} - \theta_{23}^\Lambda$ 
are almost independent of the scenario, but depend on the
value of $\alpha_3^\Lambda$. This may be shown as follows:
at large values of $m_0^\Lambda$ that are needed to
have significant RG running, at the high scale  
$|\rho_A^\Lambda| \ll 1$, so that $\Gamma^\Lambda \approx 1/\rho_A^\Lambda$,
and the value of $\alpha_2^\Lambda$ is restricted 
to be very close to $\pi$.
As a result, (\ref{kij-rhoeps}) gives
\beq
\Delta \theta_{23} \approx \frac{\Delta_\tau}{\rho_A^\Lambda}
\sin 2\theta_{23}^\Lambda (1 - \cos 2\theta_{12}^\Lambda 
\cos \alpha_3^\Lambda)
\eeq
which is almost independent of the scenario on account of
similar values of $\theta_{23}^\Lambda$ and 
$\theta_{12}^\Lambda$, but may vary by 
$\approx \cos 2\theta_{12}^\Lambda \approx 30\%$ depending
on the value of $\alpha_3^\Lambda$. 
As a consequence, a measurement of $\theta_{23}$
cannot discriminate
between the four scenarios unless it is accurate to the
level of a degree. 
However, if a scenario has already been identified, the measured 
value of $\theta_{23}$ will restrict the allowed values of the
Majorana phase $\alpha_3^\Lambda$.

\section{Discussion and summary}
\label{concl}

The current data on neutrino masses and mixings angles are
consistent with
symmetry-based schemes like quark-lepton complementarity (QLC) or
tribimaximal mixing (TBM). These scenarios predict specific
values of the neutrino mixing angles, which need to be compared
with their forthcoming more accurately measured values 
in order to confirm 
or exclude a particular postulated symmetry pattern.
However, the symmetry relations need to be imposed at a high
scale, e.g. the seesaw scale $\sim 10^{12}$ GeV, where the
neutrino masses originate.
Radiative corrections to the neutrino masses in general do not
respect the symmetries involved in QLC or TBM. As a result,
predictions of the neutrino parameters measured at 
laboratory energies become different from those given
by these symmetries at the high scale. 
It is therefore necessary to obtain low scale predictions
of these scenarios.

We have calculated radiative corrections to the neutrino 
mixing angles $\theta_{ij}$ in the context of the 
minimal supersymmetric standard model (MSSM).
We have taken the low scale to be the supersymmetry breaking
scale $\sim 10^3$ GeV and have neglected threshold
effects, if any, during this renormalisation group 
(RG) evolution.
We have presented a technique to calculate the deviations 
$\Delta\theta_{ij}$ of the mixing angles from their high scale values.
This technique yields analytically transparent results where the
errors, caused by the approximations made, are small and
under control.
The analytic treatment clarifies the dependence of the
RG running of neutrino mixing angles on
currently unknown parameters like the mass scale 
$m_0$ of neutrinos, the value of $\tan \beta$ in MSSM and 
the values of the Majorana phases.
We have also solved the RG equations
numerically to confirm that the results are indeed closely
approximated by our analytical expressions.

We have pointed out certain important patterns in the
RG evolution of the mixing angles that are valid in any
scenario. The RG running of $\theta_{12}$ always increases 
its value from the high to the low scale. Therefore, if 
a scenario predicts the value of $\theta_{12}^\Lambda$ at
the high scale $\Lambda$, the low scale measurement must 
be $\theta_{12} > \theta_{12}^\Lambda$ in order for the 
scenario to stay valid.
Similarly, the value of $\theta_{23}$ increases (decreases),
while running from a high to a low scale, for a normal
(inverted) neutrino mass ordering.
The value of $\theta_{13}$ is controlled not only by
mass ordering, but also by the values of the Majorana phases,
depending on which it may increase or decrease with
RG running.

We have considered two versions of the QLC principle (QLC1 and QLC2)
and two versions of the TBM scheme (TBM1 and TBM2), 
whose predictions at the high scale are consistent with the 
measured neutrino mixing angles at laboratory energies.
The PMNS mixing matrices at the high scale predicted
within these scenarios are 
$V_{\rm CKM}^\dagger U_{\nu, {\rm bm}}$ (QLC1),
$U_{\nu, {\rm bm}} V_{\rm CKM}^\dagger$ (QLC2),
$U_{\nu {\rm tbm}}$ (TBM1) and 
$V_{\ell L}^\dagger U_{\nu, {\rm tbm}}$ (TBM2)
respectively, where 
$V_{\rm CKM}$ is the CKM matrix,
$U_{\nu, {\rm bm}}$ the bimaximal mixing matrix,
$U_{\nu, {\rm tbm}}$ the tribimaximal \cite{tbm}
mixing matrix, and $V_{\ell L}$ a 
charged lepton mixing matrix inspired by the
Georgi-Jarlskog relation \cite{GJ} at the GUT scale.

We summarise our findings in three items:

(i) The RG running of the mixing angles should not be
too large lest the low energy values differ too much
from their high energy predictions.
Since $\theta_{12}$ is the most accurately measured angle
currently, that puts strong constraints on
the allowed values of $m_0^\Lambda \tan\beta$ as well as on 
the Majorana phase $\alpha_2^\Lambda$. 
It is observed that, for $m_0^\Lambda \tan\beta \gsim 2$ eV
(i.e. $m_0 \tan\beta \gsim 1.4$ eV),
the allowed range of $\alpha_2^\Lambda$ is severely restricted.
In all the scenarios, $m_0^\Lambda \tan\beta$ has to be
$\lsim 4.4$ eV (i.e.  $m_0 \tan\beta$ has to be
$\lsim 3.1$ eV)  for consistency with data. 
Moreover, the larger the value of $m_0 \tan\beta$, 
the closer to $\pi$ has to be the value of $\alpha_2^\Lambda$,
and hence that of $\alpha_2$ at the laboratory scale,
cf. eq.~(\ref{a2-expr}).
Thus we find a preference for the approximate equality
$m_1 \simeq - m_2$ (especially for large $\tan\beta$),
as also suggested by considerations of leptogenesis
\cite{leptogenesis}. 
A  reduction of errors on the $\theta_{12}$ measurement would
decrease the allowed area in the $m_0^\Lambda \tan\beta$ 
-- $\alpha_2^\Lambda$ plane.
Moreover, since $\theta_{12}$ always increases from higher to lower scales,
if its value is measured to be
smaller than what is predicted at the high scale
in a scenario, that
particular scenario would get excluded. 
{\it All the scenarios 
considered in this paper would be excluded if $\theta_{12}$
were measured to be less than $32^\circ$.}

(ii) The measurement of $\theta_{13}$ is most likely
to serve as a discriminator among the four scenarios
considered here. The predicted values of $\theta_{13}$ in these scenarios 
at the high scale are $8.9^\circ$ (QLC1), $1.6^\circ$ (QLC2), 
$0^\circ$ (TBM1) and $3.1^\circ$ (TBM2).
RG running can modify the value of $\theta_{13}$ in either
direction; however, the restrictions on $m_0^\Lambda \tan\beta$ and
$\alpha_2^\Lambda$ from the $\theta_{12}$ measurements limit
the extent of this modification. We find, for example, that
the value of $\theta_{13}$ in QLC1 cannot be less than $6^\circ$
($3 \sigma$), whereas in none of the other cases can 
$\theta_{13}$ be as large as $6^\circ$ within $3\sigma$. 
Neutrino experiments during
the next decade should be able to measure the value of 
$\theta_{13}$ if it is greater than $\approx 5^\circ$
or to put an upper bound of $\approx 5^\circ$ on it. In either
case, the scanario QLC1 will be distinguishable from the
others. 
Both QLC2 and TBM1 predict almost identical $\theta_{13}$
ranges: $\theta_{13} < 3^\circ$ ($3\sigma$). 
The allowed $3\sigma$ range of $\theta_{13}$ for TBM2 
overlaps with the QLC2/TBM1 range for 
$m_0^\Lambda \tan\beta \gsim 2$ eV
(i.e. $m_0 \tan\beta \gsim 1.4$ eV).
Limiting $m_0 \tan\beta$ to lesser values
would also help in discriminating between TBM2 on one hand,
and QLC1/TBM2 on the other. 

(iii) It is not possible for a $\theta_{23}$ measurement to 
discriminate among the four scenarios unless it is 
accurate to the level of a degree. 
However the value of $\theta_{23}$ within any scenario 
is strongly dependent on the Majorana phase $\alpha_3^\Lambda$.
Therefore, if a scenario has already been identified, the measured 
value of $\theta_{23}$ will restrict the allowed values of 
$\alpha_3^\Lambda$.

In conclusion, we have shown how the high scale predictions
on neutrino mixing angles get modified with RG running in
MSSM for four symmetry-inspired scenarios that are consistent
with the current neutrino data.
With a combination of analytical insights and numerical 
calculations, we show that this limits the allowed ranges
of parameters like $m_0^\Lambda$, $\tan\beta$ and the Majorana phases.
We also indicate the extent to which future measurements can 
discriminate among various scenarios and how the values of
the parameters may be further restricted.

\section*{Acknowledgements}

A.D. and S.G. would like to thank W. Rodejohann 
for useful discussions.
P.R. acknowledges the hospitality of the University of Hawaii 
at Manoa where part of this work was carried out.
The work of A.D. is partly supported through the 
Partner Group program between the Max Planck Institute
for Physics and Tata Institute of Fundamental Research.
Part of the computational work for this study were carried 
out at cluster computing facility in the Harish-Chandra 
Research Institute (http://www.cluster.mri.ernet.in).


\end{document}